\documentclass[dvips]{article}

\usepackage[numaddn=\pi,per=fraction]{siunitx}
\usepackage{booktabs, tabularx}

\usepackage[switch]{lineno}
\usepackage{icrc2011}

\title{
  Mass Composition Sensitivity of an Array of Water Cherenkov and
  Scintillation Detectors
}

\newcommand{\etal}{\MakeLowercase{\textit{et al. }}} 
\shorttitle{Javier G. Gonzalez \etal Scintillation Detector Array}

\authors{Javier G. Gonzalez, Ralph Engel, Markus Roth }
\afiliations{
  Karlsruhe Institute of Technology (KIT), Karlsruhe, Germany
}
\email{javier.gonzalez@ik.fzk.de}

\abstract{ We consider a hybrid array composed of
scintillation and water Cherenkov detectors designed to measure the
cosmic ray primary mass composition at energies of about 1 EeV. We have
developed a simulation and reconstruction chain to study the
theoretical performance of such an array. In this work we investigate
the sensitivity of mass composition observables in relation to the
geometry of the array. The detectors are arranged in a triangular grid
with fixed 750 m spacing and the configuration of the scintillator
detectors is optimized for mass composition sensitivity. We show that
the performance for composition determination can be compared favorably
to that of $X_{\text{max}}$ measurements after the difference in duty
cycles is considered.
}

\begin{document}
\maketitle

\section{Introduction}

The measurement of the mass-composition of ultra-high energy cosmic
rays is one of the keys that can help us elucidate their origin. Up
to energies around \SI{e15}{eV}, the Galaxy is believed to be the
source of cosmic rays. Several acceleration mechanisms are certainly
at play but it is widely expected that the dominant one is first order
Fermi acceleration at the vicinity of supernova remnant shock
waves. These Galactic accelerators should
theoretically become inefficient between
\SI{e15}{eV} and \SI{e18}{eV}. The KASCADE experiment has
measured the energy spectra for different mass groups in this energy
range and found that there is a steepening of the spectra at an energy
that increases with the cosmic ray mass \cite{Apel:2008cd}. As a
result, the mass-composition becomes progressively heavy. It is also
thought that extra-galactic sources can start to contribute to the
total cosmic ray flux at energies above \SI{4e17}{eV}. The
onset of such an extra-galactic component would probably produce
another change in composition.
The $X_{\text{max}}$ measurements from the HiRes-MIA experiment have
been interpreted as a change in composition, from heavy to light,
starting at \SI{4e17}{eV} and becoming proton-dominated at
\SI{1.6e18}{eV} \cite{AbuZayyad:2000ay, Abbasi:2009nf} while the
measurements from the Pierre Auger observatory hint at a light or mixed composition
that becomes heavier beyond \SI{2e18}{eV} \cite{Unger:2009kk}.

\subsection{Measuring Composition}

Roughly speaking, the techniques for inferring the mass composition of
cosmic rays can be split in two categories, depending on whether they exploit the
sensitivity to the depth of shower maximum ($X_{\text{max}}$) or to the ratio of the
muon and electromagnetic components of the air shower
\cite{Nagano:2000ve}. Direct measurements of the fluorescence
emission fall in the first category, and so do the various
measurements of the Cherenkov light produced by air showers. Most ground-based
detector observables depend one way or another on the number of
muons in the air shower. However, the arrival time profile of shower
particles has been used as an observable mostly sensitive to
$X_{\text{max}}$, in particular the so-called \textit{rise-time}, the time it
takes for the signal to rise from 10\% to 50\% of the integrated
signal \cite{Wahlberg:2009zz}.
The measurement of the number of muons and electrons in the air shower
can be done directly, for example, the way it was done with the
KASCADE detector \cite{Antoni:2003gd}.

The Pierre Auger Observatory is developing a series of enhancements
that aim at the energy range between \SI{e17}{eV} and \SI{e19}{eV}
\cite{amiga,heat}. In particular, the objective of the AMIGA
enhancement \cite{amiga} is to measure the muon component of the air shower using
scintillators shielded by several meters of soil. In the same spirit, we are considering a
\textit{hybrid} surface array, consisting of two super-imposed ground
arrays, a Water Cherenkov Detector (WCD) array and a scintillation detector array. The purpose of
the scintillation detectors is to increase the sensitivity to the
electromagnetic part of the air shower.

We will consider an array of stations arranged in a triangular grid
with a separation of \SI{750}{m}. Each scintillator station is made
with \SI{3}{cm} thick plastic scintillator tiles like the ones used
for the muon detectors in the KASCADE array.  In order to enhance the
signal from gamma rays in air showers, we study the effect of adding a
certain amount of lead on top of the scintillators. For
conversion of around 80\% of the high energy gamma rays one normally
needs a shielding of about 2 radiation lengths. One radiation length
corresponds to \SI{0.56}{cm} of lead.

\subsection{Hybrid Surface Detector Performance}

\begin{figure*}[!t]
  \centerline{
    \includegraphics[width=0.5\linewidth]{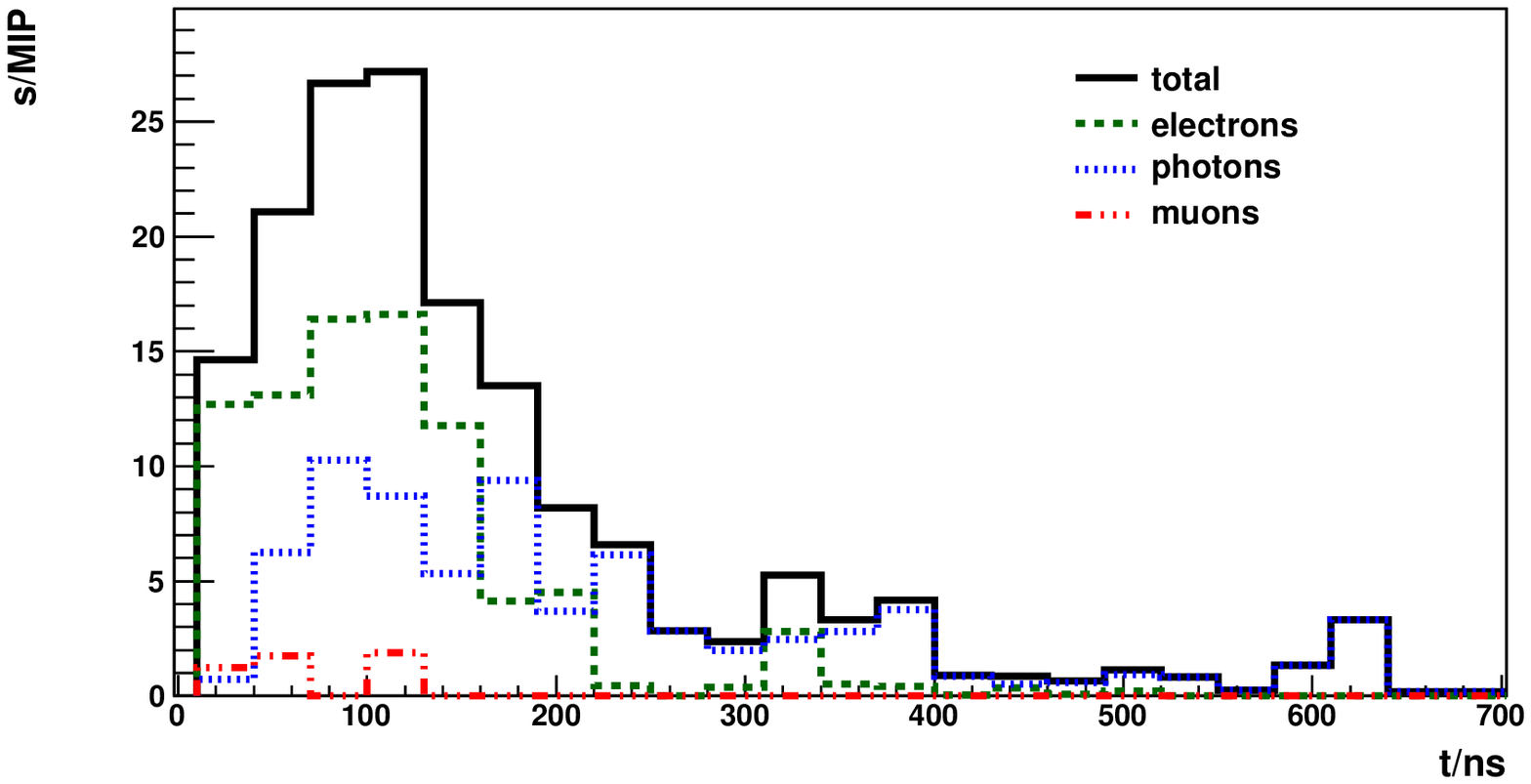}
    \hfil
    \includegraphics[width=0.5\linewidth]{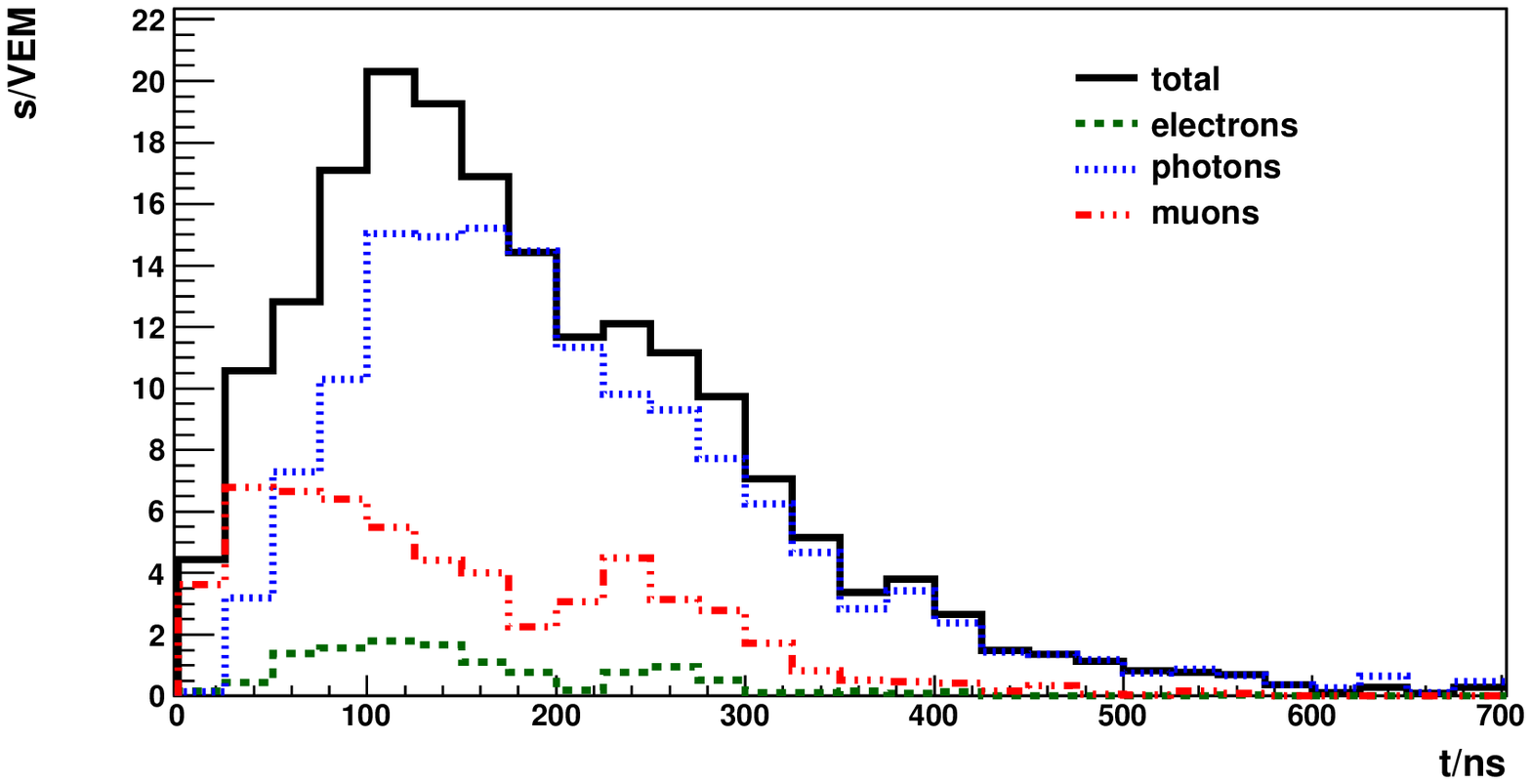}
  }
  \caption{\small Comparison of WCD and scintillator traces. These two sets of traces correspond to a pair of scintillator station with no photon converter (left) and a WCD station (right). They are located at \SI{413}{m} from the axis of a \SI{e18}{eV} shower arriving at \SI{38}{\degree}. Note that, while the WCD station already has a significant contribution from muons, this is very small in the scintillator, which is dominated by the signal from electrons.}
  \label{fig:traces}
\end{figure*}

In order to study such a hybrid detector, we have implemented a
simulation and reconstruction chain based on the Pierre Auger
Observatory offline framework \cite{Argiro:2007qg}. Using this chain, we have
processed 20,000 simulated air showers produced with the CORSIKA
\cite{corsika} event generator. All these showers were generated using
QGSJET II for high energy hadronic interaction simulations
\cite{Ostapchenko:2010vb}.

The parameter space chosen for the simulations has been determined by
the energy and angular ranges under consideration. Additionally, it has
been shown recently that real air-showers appear to have a larger muon
component than simulated air-showers \cite{muon_excess}. For this
reason we have also performed the simulations with the number of muons
artificially increased by a factor of two, which gives the observed
muon contribution after an energy shift of about 28\%.  We have also
investigated the effect of placing a certain amount of shielding
on top of each scintillator station, as well as the effect of changing its
area. In total there are 600 parameter combinations, with 100
simulated showers each, for a total of 60,000 simulated events.
The parameters are shown in table \ref{table:parameters}.

\begin{table}[!h]
\begin{center}
\begin{tabular}{lc}
\toprule
Primaries & p, Fe \\
E (eV)  & \SI{e17.75}{},  \SI{e18}{},   \SI{e18.25}{}     \\
$\theta$  & \SI{0}{\degree},  \SI{25}{\degree},  \SI{38}{\degree},  \SI{48}{\degree},  \SI{60}{\degree} \\
Area  & \SI{3}{m^2},  \SI{10}{m^2} \\
$\mu_{scale}$ & 1,  2  \\
Shield (rad. lengths)& 0, 0.25, 0.5, 1, 2 \\
\midrule
\end{tabular}
\caption{Parameters considered in the simulation of the response of
  the scintillator array.}\label{table:parameters}
\end{center}
\end{table}

The simulation of the interactions of the shower particles with the
detector is done using the Geant4 package \cite{geant4, geant4_2}. The
scintillation efficiencies used in the simulation correspond to the
specifications for Bicron's BD-416 scintillators: Polyvinyl
Toluene scintillators with a nominal light yield of about
\SI{e4}{photons/MeV} and a density of \SI{1.032}{g/cm^3}. The
resulting scintillation photons are sampled at a frequency of
\SI{100}{MHz} to produce one FADC trace per station. The signal in
each station is measured in units of \textit{Minimum Ionizing Particle}
equivalent, or \textit{MIP}, where a MIP is given by the position of
the peak of the Landau distribution for vertical muons. We then
consider only stations with signals between 1 and \SI{2000}{MIP} in
order to simulate the dynamical range.

The arrival direction and core
position of each event are currently estimated using only the WCDs. The arrival
direction is determined by fitting a spherical shower front to the
signal start times of the stations in the event. The core position is
determined by adjusting a lateral distribution function (LDF) of the
form
\begin{equation}
  S_{\text{WCD}}(r)=S_{r_0}\,\left(\frac{r}{r_0}\right)^{-\beta}
\end{equation}
to the total signal in the stations in the event. The $r_0$ parameter
is \SI{450}{m} and $S_{450}$ is the usual energy estimator for a WCD
array with this grid spacing because it is the optimum distance for
determining the signal \cite{Newton:2006wy}.
The LDF for the scintillator array is described by the same formula
and the optimum $r_0$ that will enhance the composition sensitivity
still needs to be determined.

\begin{figure}[!t]
  \begin{center}
    \includegraphics[width=\linewidth]{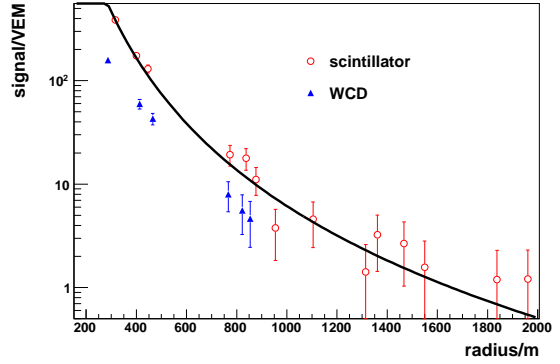}
  \end{center}
  \caption{\small A typical \SI{e18}{eV} shower arriving at
    \SI{38}{\degree} showing the reconstructed LDF of the scintillator array.}
  \label{fig:sample_event}
\end{figure}

\begin{figure}[!t]
  \begin{center}
    \includegraphics[width=\linewidth]{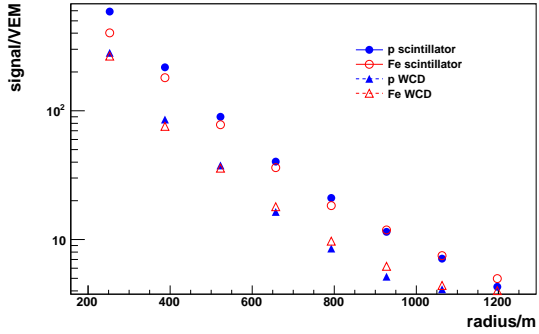}
  \end{center}
  \caption{\small Comparison of the lateral distributions of \SI{1e18}{eV} proton
    and iron induced air showers arriving at \SI{38}{degree} using a combined scintillator/WCD array. The area of each station is \SI{10}{m^2}. The muon component is artificially scaled by a factor of two.}
  \label{fig:ldf_comparison}
\end{figure}

One typical event is
displayed in figures \ref{fig:traces} and \ref{fig:sample_event}.
An example of the resulting average lateral distribution functions
over 100 showers, for both the scintillator and the WCD array,
can be seen in figure \ref{fig:ldf_comparison}.
Clearly, the slope of
the proton and iron LDFs differ and there is a cross-over point where
the signals are equal.  This cross-over point depends on energy and
zenith angle and this dependence is different for the two arrays. At
\SI{e18}{eV} and \SI{0}{\degree}, it is close to the shower axis for
both arrays. As the zenith angle increases, the cross-over occurs at
greater distances to the shower axis and for angles greater than
\SI{25}{\degree}, which correspond to most of the aperture, it is
located at more than 600 meters from the axis. At larger distances,
the differences between proton and iron signals are similar in the
scintillators and the WCDs since they both detect mostly muons in this
radial range. Therefore, the largest difference in signal between
proton and iron is found close to the axis.

\begin{figure}[t]
  \begin{center}
    \includegraphics[width=\linewidth]{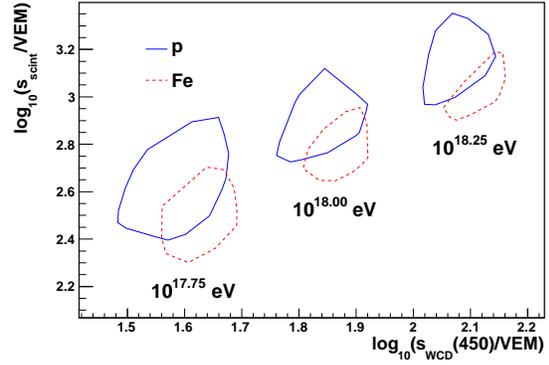}
  \end{center}
  \caption{\small 68\% contours of the $\log_{10}(S_{\text{sci}})-\log_{10}(S_{\text{WCD}})$
    scatter plot. This corresponds to a zenith angle of
    \SI{25}{\degree}, \SI{10}{m^2} stations with no lead photon converter and
    a scaling of the muon component by a factor of two.
  }
  \label{fig:s_sci_s_wcd_scatter}
\end{figure}

For the reason mentioned above, we use the signal at 400 meters from
the shower axis as the measure for the scintillator array signal. For
each event we can then correlate the scintillator array signal
($S_{\text{sci}}$) with the WCD signal at 450 m ($S_{\text{WCD}}$).  A typical contour
plot of these quantities is shown in figure
\ref{fig:s_sci_s_wcd_scatter}. One can model the dependence of the
scintillator and WCD signals as a function of the energy and the mass
number using

\begin{equation}
  \log_{10}S_i = a_i + b_i\log_{10}E + c_i\log_{10}A
  \label{eq:signal_vs_e_a}
\end{equation}

This set of equations can then be inverted to find $\log_{10}E$ and
$\log_{10}A$.
The result is a mass
estimator and an unbiased energy estimator.

There are various ways of measuring the discrimination power of a
given statistic. One of them is the figure of merit:
\begin{equation}
  f = \frac{\langle s_{Fe} \rangle - \langle s_{p}
    \rangle}{\sqrt{\sigma_{Fe}^2 + \sigma_{p}^2}}.
  \label{eq:merit_figure}
\end{equation}

\begin{figure}[t]
  \begin{center}
    \includegraphics[width=\linewidth]{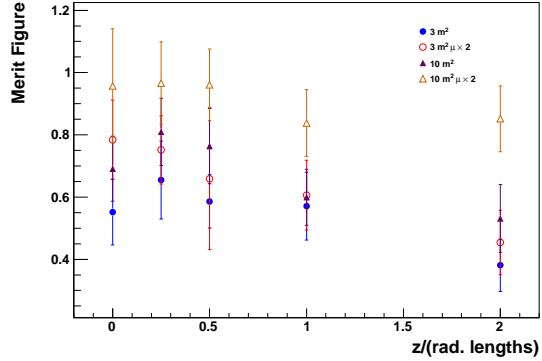}
  \end{center}
  \caption{\small Average figure of merit as a function of the
    thickness of the lead shield. The different graphs correspond to the
    different detector areas and muon component scaling factor considered.}
  \label{fig:ave_discrimination}
\end{figure}

The resulting average discrimination power, after correcting by the
decrease in aperture with increasing zenith angle, is displayed in
figure \ref{fig:ave_discrimination}.  We have used other multivariate
methods as well as alternative measures of discrimination power,
such as the efficiency at given purity levels. The results are
qualitatively the same.

As an exercise to get an idea of the magnitude of the resulting
uncertainties in the determination of the composition, one can pose
the problem of determining the fraction of protons in a sample of
proton and iron events.

Given a sample of equal number proton and iron showers, we can
determine a cut value for the mass estimator such that the error rates
of the first and second kind are equal ($\epsilon$). Let's now consider a sample of N events where a fraction $f$ of them are protons. The observed
number of {\textit proton-like} ($s_{obs}$, the number of events passing the cut) and {\textit
  iron-like} ($b_{obs}$) events depend on the number of
actual signal and background events through the following equation:
\begin{equation}
\left(
\begin{array}{c}
s_{obs} \\
b_{obs}
\end{array}
\right)
=
\left(
\begin{array}{cc}
1-\epsilon & \epsilon \\
\epsilon & 1-\epsilon
\end{array}
\right)
\left(
\begin{array}{c}
fN \\
(1-f)N
\end{array}
\right)
\end{equation}
and this equation can be inverted to estimate the proton fraction.

\begin{figure}[t]
  \begin{center}
    \includegraphics[width=\linewidth]{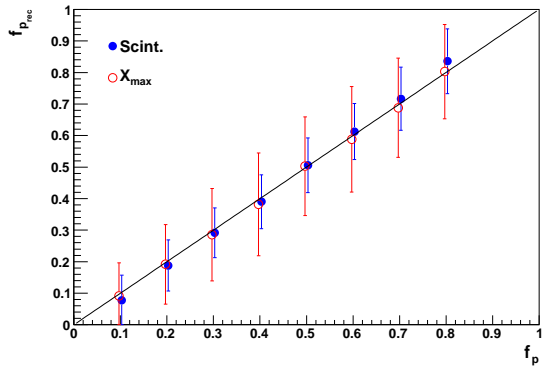}
  \end{center}
  \caption{\small Cross validation of the proton fraction
    determination.  This corresponds to \SI{e18} eV showers arriving
    at \SI{38}{\degree} and muon component scaled by a factor of
    two. The stations have an area of \SI{3}{m^2} and 0.25 radiation length photon
    converters. The test sample for the $X_{\text{max}}$ technique is
    30\% of the sample for the combined array.}
  \label{fig:rec_fraction}
\end{figure}

In order to judge the quality of the method, a common strategy is to
split the sample into a training sample, from which we get the values
of the coefficients in equations \ref{eq:signal_vs_e_a} and the value
of the cut. We can also use a k-fold cross validation method. This
consists in splitting the sample in sets of k events, taking half of
the sets to form the training sample and use the rest as a test
sample. This process is then repeated with different combinations of
sets. The same procedure is repeated using the distributions of
$X_{\text{max}}$ for proton and iron but in this case we limit the size
of the test sample to 30\% in order to take the combined effect of the
aperture and limited duty cycle of a fluorescence detector into
account. The result can be seen in figure \ref{fig:rec_fraction}.

From this simple exercise we have learned that both methods give
similar uncertainties in the determination of the proton fraction. In
other words, that the combined scintillator/WCD array provides a
composition estimate that is competitive with that from fluorescence
detectors after one considers the fluorescence detectors' limited
duty-cycle. We are aware that the precise factor to penalize the
$X_{\text{max}}$ measurement will depend on the relative apertures of
the fluorescence detector and the surface detector and that there will
be systematic errors deriving from the different strategies in
training the classification and regression methods. This number will
still give the right order of magnitude.

\section{Summary}

We have performed detailed simulations of the response of a combined
scintillator/WCD array to air showers produced by cosmic ray with primary
energies around \SI{1e18}{eV}. We have considered different
configurations for the individual scintillator stations in order to
optimize the composition sensitivity of the combined array.

We have found that an array of \SI{10}{m^2} scintillator stations,
together with the WCD array, can provide an estimate of the average
mass composition that is competitive with $X_{\text{max}}$
measurements after differences in duty-cycle are considered and
that the addition of photon converters on top of each station
does not add to the separation power.

In this work we only considered a grid spacing of \SI{750}{m} but we
have noted that the highest separation between proton and iron occurs
at small distances to the core. Better knowledge of the impact of
smaller spacing and of the precise optimum distance at which to
measure the scintillator signal requires further study.


\clearpage

\end{document}